\documentclass[a4paper]{article}

\usepackage{INTERSPEECH2018, multirow}

\title{Simultaneous Denoising and Dereverberation Using Deep Embedding Features}
\name{Cunhang Fan$^{1,2}$, Jianhua Tao$^{1,2,3}$, Bin Liu$^1$, Jiangyan Yi$^1$, Zhengqi Wen$^1$,}
\address{
	$^1$National Laboratory of Pattern Recognition, Institute of Automation, Chinese Academy of Sciences, Beijing, China\\
	$^2$School of Artificial Intelligence, University of Chinese Academy of Sciences, Beijing, China\\
	$^3$CAS Center for Excellence in Brain Science and Intelligence Technology, Beijing, China}
\email{\{cunhang.fan, jhtao, liubin, jiangyan.yi, zqwen\}\@nlpr.ia.ac.cn}

\begin{document}

\maketitle
\begin{abstract}
  Monaural speech dereverberation is a very challenging task because no spatial cues can be used. When the additive noises exist, this task becomes more challenging. In this paper, we propose a joint training method for simultaneous speech denoising and dereverberation using deep embedding features, which is based on the deep clustering (DC). DC is a state-of-the-art method for speech separation that includes embedding learning and K-means clustering. As for our proposed method, it contains two stages: denoising and dereverberation. At the denoising stage, the DC network is leveraged to extract noise-free deep embedding features. These embedding features are generated from the anechoic speech and residual reverberation signals. They can represent the inferred spectral masking patterns of the desired signals, which are discriminative features. At the dereverberation stage, instead of using the unsupervised K-means clustering algorithm, another supervised neural network is utilized to estimate the anechoic speech from these deep embedding features. Finally, the denoising stage and dereverberation stage are optimized by the joint training method. Experimental results show that the proposed method outperforms the WPE and BLSTM baselines, especially in the low SNR condition.
\end{abstract}

\section{Introduction}

In many real-world speech communication applications, speech signals recorded by receivers not only contain the desired speech signals, but also the reverberation and additive noises. However, these reverberations and noises can degrade speech intelligibility and sound quality for human listeners \cite{helfer1990hearing,beutelmann2006prediction,warzybok2014subjective,fan2019noise}. In this paper, we focus on the single-channel simultaneous speech denoising and dereverberation.

In order to address the speech dereverberation problem, many algorithms have been proposed in the past decades \cite{kodrasi2018joint,nakatani2010speech,yoshioka2012generalization}. Weighted prediction error (WPE) \cite{nakatani2010speech,yoshioka2012generalization} methods deal with reverberation on a signal level, which have been suggested to be very efficient at suppressing room acoustic effects. They are based on delayed linear prediction. WPE first obtains the frequency-dependent linear prediction filters by a number of history frames. Then the enhanced signal is acquired by subtracting the filtered signal from the original reverberant signal in the subband domain. These algorithms can reduce reverberation well in clean condition. However, when there are the additive noises and reverberation simultaneity, the performance of these algorithms are suffered severely.

In recent years, deep neural networks (DNNs) have emerged as a powerful learning method and have been gradually applied to speech dereverberation \cite{han2015learning,wu2017reverberation,williamson2017time,zhao2018late}. In \cite{han2015learning}, Han et al. propose to use DNN to learn a spectral mapping from reverberation to anechoic speech. While at low reverberation time (RT60) the performance is still limited. And direct spectral magnitude estimation performs worse than mask estimation \cite{wang2014training}. Williamson and Wang \cite{williamson2017time} propose to do the speech dereverberation in the complex domain, they use a DNN to estimate a complex ideal ratio mask (cIRM) rather than the spectral magnitude. And the magnitude and phase spectrum are enhanced jointly. Although this method can get a better performance than \cite{han2015learning}, the computational cost is too expensive.

In our previous work, we proposed a deep embedding features method for speech separation \cite{fan2019discriminitive,fan2020end,fan2020deep}, which was based on deep clustering (DC) \cite{Hershey2016Deep}. These deep embedding features can be regarded as very discriminative features for speech dereverberation, which can discriminate the anechoic speech and the reverberant signals very well. Motived by this, in this study, we propose a joint training method for simultaneous speech denoising and dereverberation using deep embedding features. The proposed method includes two stages: denoising and dereverberation. Firstly, a DC network is trained to extract deep embedding features without noise signals, which is the denoising stage. These embedding features are generated from the anechoic speech and residual reverberation signals. They are noise-free vectors. In addition, they can represent the inferred spectral masking patterns of the desired signals, which have an advantage in discriminating anechoic and reverberation. Secondly, at the dereverberation stage, instead of using the unsupervised K-means clustering algorithm, another supervised neural network is applied to learn the mask of anechoic speech from these deep embedding features. In this stage, the objective function can be directly defined at the desired signals not the embedding vectors, which is conducive to dereverberation. Finally, the denoising stage and dereverberation stage are optimized by the joint training method. In this way, noise reduction and dereverberation can be simultaneously optimized so that the performance of speech enhancement can be improved.

The rest of this paper is organized as follows. Section 2 presents the signal channel speech dereverberation based on mask. The  proposed method is stated in section 3. Section 4 shows detailed experiments and results. Section 5 draws conclusions.

\section{Single Channel simultaneous denoising and dereverberation method Based on Mask}


Let \(x(t)\) and \(h(t)\) denote anechoic speech and room impulse response (RIR), respectively. \(n(t)\) represents the additive noise. The noisy and reverberant speech \(y(t)\) can be represented as:
\begin{equation}
y(t) = x(t)*h(t)+n(t)
\label{eq1}
\end{equation}
After the short-time Fourier transformation (STFT), this following relationship is still satisfied:
\begin{equation}
Y(t,f) = X(t,f)\times{H}+N(t,f)
\label{eq2}
\end{equation}
where \(Y(t,f)\), \(X(t,f)\) and \(N(t,f)\) denote the STFT of \(y(t)\), \(x(t)\) and \(n(t)\), respectively.

The objective of this study is to estimate the clean anechoic speech from \(y(t)\) or \(Y(t,f)\). As for speech separation task, it is well known that mask based speech separation can obtain a better result \cite{fan2019discriminitive,fan2018Utterance,wang2018supervised,fan2020spatial}. Similarly, we apply the mask \(M(t,f)\) for speech dereverberation in this paper. According to the commonly used masking method, the estimated magnitude  \(|\widetilde{X}(t,f)|\)  of anechoic can be estimated by
%
\begin{equation}
|\widetilde{X}(t,f)|=|Y(t,f) |\odot{M(t,f)}
\label{eq3}
\end{equation}
where \(\odot\) indicates element-wise multiplication. Finally, the estimated magnitude  \(|\widetilde{X}(t,f)|\) and the phase of noisy reverberant signal are used to reconstruct anechoic speech by inverse STFT (ISTFT). 

\section{Our proposed method}

\begin{figure}[t]
	\centering
	\includegraphics[width=\linewidth]{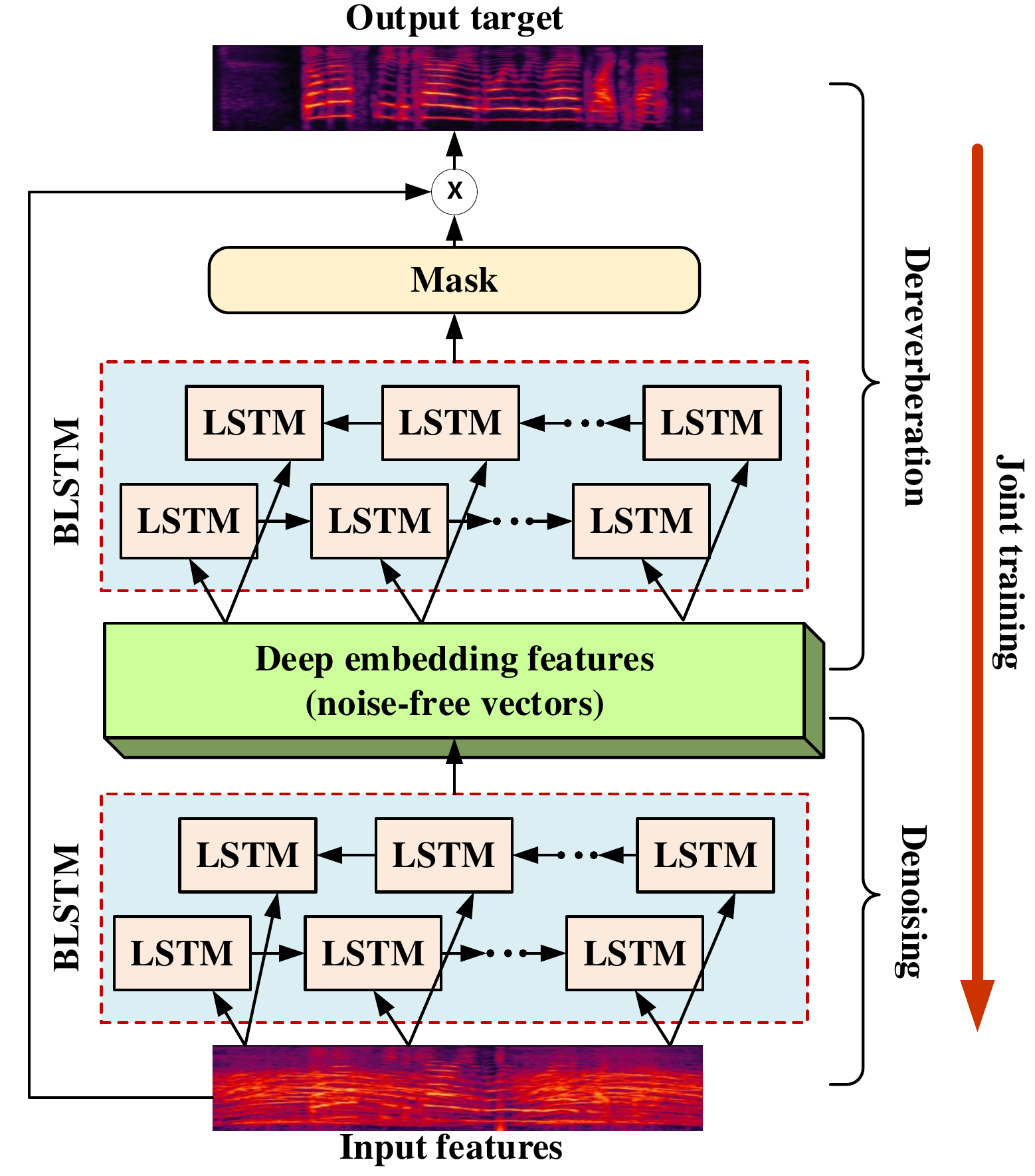}
	\caption{Schematic diagram of our proposed speech dereverberation system with deep clustering.}
	\label{fig:DC_SD}
\end{figure}

In this paper, we extend our previous work\cite{fan2019discriminitive} to simultaneous speech denoising and dereverberation as shown in Fig.~\ref{fig:DC_SD}. It includes two stages: speech denoising and speech dereverberation. Firstly, at the speech denoising stage, we utilize the DC network to extract the deep embedding features, which are noise-free vectors. Secondly, at the speech dereverberation stage, instead of using the unsupervised K-means clustering algorithm, another supervised neural network is applied to learn the mask of the target signals from these deep embedding features. The reason is that these features can discriminate the anechoic speech and the reverberant signals very well. Finally, in order to improve the performance of the proposed system, these two stages are optimized by the joint training method.

\subsection{Speech denoising stage}

At the speech denoising stage, we firstly train a DC \cite{Hershey2016Deep} network based on BLSTM as the extractor of \(D\)-dimensional deep embedding features \(V\in{\mathbb{R}^{TF\times{D}}}\). The embedding \(V\) can be regarded as a noise-free and discriminative feature encoding of the signal partition. Here we consider a unit-norm embedding, so
\begin{equation}
|v_i|^2=1,\quad v_i={v_{i,d}}
\label{eq5}
\end{equation}
where \(v_{i,d}\) is the value of the \(d\)-th dimension of \(V\) for element \(i\). We let the embeddings \(V\) to implicitly represent an \(TF\times{TF}\) estimated affinity matrix \(VV^T\).

The loss function of deep embedding features network is defined as follow:
\begin{equation}
\begin{split}
J_{DC} &=||VV^T-BB^T||_F^2\\
&=||VV^T||_F^2-2||V^TB||_F^2+||BB^T||_F^2\\
\end{split}
\label{eq6}
\end{equation}
where \(||*||_F^2\) is the squared Frobenius norm. \(B\in{\mathbb{R}^{TF\times{2}}}\) is the target membership indicator for each T-F bin. In order to remove the noise firstly, we define the indicator \(B\) between anechoic speech and reverberant signal. Therefore, the indicator \(B\) maps each element \(tf\) to each cluster of anechoic speech or reverberant signal. If the anechoic speech has the highest energy at time \(t\) and frequency \(f\) compared to the reverberant signal, \(B_{tf,1}=1\) and \(B_{tf,2}=0\). Otherwise, \(B_{tf,1}=0\) and \(B_{tf,2}=1\). In this case, \(BB^T\) is considered as a binary affinity matrix that represents the cluster assignments.

\subsection{Speech dereverberation stage}

The dereverberation stage aims to estimate the anechoic speech from the noise-free deep embedding features. Because these embedding features can represent the inferred spectral masking patterns of the desired signals and they are discriminative. We use them as the input features of our dereverberation stage.

As for DC \cite{Hershey2016Deep}, it utilizes the supervised network to extract the deep embedding vectors. However, it uses the unsupervised K-means clustering algorithm to estimate the target mask. In this study, instead of using the unsupervised K-means, we apply another supervised BLSTM network to learn the mask of anechoic speech so that the proposed method can remove the reverberation very well. 

As shown in Fig.~\ref{fig:DC_SD}, when these deep embedding features (noise-free vectors) are extracted, they are inputted to the BLSTM network to estimate the target mask: 
\begin{equation}
\widetilde{M}(t,f)=\psi_{BLSTM}(V)
\label{eq8}
\end{equation}
where \(\widetilde{M}(t,f)\) denotes the estimated mask of anechoic speech and \(\psi_{BLSTM}(*)\) is a mapping function based on the BLSTM network.


\subsection{Joint training}

In order to improve the performance of the proposed system, the joint training method is applied to optimize the denoising stage and dereverberation stage, simultaneously. We directly apply the mean square error (MSE) between estimated magnitude and true magnitude as the training criterion. Therefore, the loss function of our proposed method is defined as the following:
\begin{equation}
J=\frac{1}{TF}\sum|||Y(t,f)|\odot{\widetilde{M}(t,f)}-|X(t,f)|||_F^2
\label{eq10}
\end{equation}


In order to get better deep embedding features, we train the DC network firstly with loss function Eq. \ref{eq6}. Then the denoising stage and dereverberation stage are optimized by the joint training with loss function Eq. \ref{eq10}.

\section{Experiments and Results}
\label{sec:typestyle}

\subsection{Dataset}
\label{ssec:subhead1}

\begin{table}[t]
	\caption{Configurations used for simulating training data.}
	\label{tab:training set}
	\centering
	\begin{tabular}{c|c}
		\toprule
		{\multirow{2}*{Dataset}} & 4620 utterances in training set from\\
		& the TIMIT training database\\
		\midrule
		\(RT_{60}\) & 0.2s to 2s with a step size of 0.2s \\
		\midrule
		Noise database & 100 Nonspeech Sounds \\
		\midrule
		SNR(dB) & -5, 0, 5, 10 \\
		\bottomrule
	\end{tabular}
\end{table}

The experiment is conducted using the TIMIT database \cite{garofolo1993darpa}, which has 630 speakers each speaking 10 utterances. We create the training, validation and test sets in the same manner. The reverberant microphone signals are generated by convolving the clean utterances with different RIRs, which is similar to \cite{kodrasi2018single}. The RIRs are generated using the image-source method \cite{RIR}. The noises use in the training and validation sets include 100 different noise types, which can be download from \cite{100nonspeech}. 

In order to generate the training dataset, 4620 clean utterances from the TIMIT training database are used. And they are convolved with 10 RIRs, resulting in 46200 training utterances in total. The development dataset is generated using 551 utterances from the TIMIT training set database and 9 RIRs, resulting in 4959 utterances in total. Then the training and development dataset are mixed with 100 Nonspeech Sounds database\cite{100nonspeech} at 4 signal-to-noise ratio (SNR) (-5, 0, 5 and 10dB). Detailed configuration is listed in Table~\ref{tab:training set} and Table~\ref{tab:development set}. Finally, the testing dataset is generated using 268 clean utterances from the TIMIT testing database and 18 RIRs, resulting in 4824 utterances in total. As for the test set, besides the 100 different seen noise types, twelve unseen noises are used, which are from NISEX-92 dataset \cite{varga1993assessment}. Same to the training set, these 4824 utterances are mixed with these noises at 4 SNR (-5, 0, 5 and 10 dB). Detailed configuration is listed in Table~\ref{tab:test set}.


\begin{table}[t]
	\caption{Configurations used for simulating development data.}
	\label{tab:development set}
	\centering
	\begin{tabular}{c|c}
		\toprule
		{\multirow{2}*{Dataset}} & 551 utterances in training set from\\
		& the TIMIT training database\\
		\midrule
		\(RT_{60}\) & 0.3s to 1.9s with a step size of 0.2s \\
		\midrule
		Noise database & 100 Nonspeech Sounds \\
		\midrule
		SNR(dB) & -5, 0, 5, 10 \\
		\bottomrule
	\end{tabular}
\end{table}

\begin{table}[t]
	\caption{Configurations used for simulating test data.}
	\label{tab:test set}
	\centering
	\begin{tabular}{c|c}
		\toprule
		{\multirow{2}*{Dataset}} & 268 utterances in training set from\\
		&the TIMIT training database\\
		\midrule
		\(RT_{60}\) & 0.35s to 1.95s with a step size of 0.1s \\
		\midrule
		Noise database & 100 Nonspeech Sounds and NISEX-92 \\
		\midrule
		SNR(dB) & -5, 0, 5, 10 \\
		\bottomrule
	\end{tabular}
\end{table}

\subsection{Experimental setups}

The sampling rate of all generated data is 8 kHz before processing to reduce computational and memory costs. The 129-dim spectral magnitudes of the noisy speech are used as the input features, which are computed using a STFT with 32 ms length hamming window and 16 ms window shift. Our models are implemented using Tensorflow deep learning framework \cite{Abadi2016TensorFlow}.

In this work, the deep embedding network has two BLSTM layers with 512 units. In order to acquire a better performance, we select four different numbers of embedding dimension D (10, 20, 30 and 40). A tanh activation function is followed by the embedding layer. As for the speech dereverberation stage, it has only one BLSTM layer with 512 units. Therefore, there are three BLSTM layers for our proposed method. A Rectified Liner Uint (ReLU) activation function is followed by the dereverberation stage, which is the mask estimation layer. 

All models contain random dropouts with a dropout rate 0.5. Each minibatch contains 20 randomly selected utterances. The minimum number of epoch is 30. The learning rate is initialized as 0.0005 and scaled down by 0.7 when the training objective function value increased on the development set. Our models are optimized with the Adam algorithm \cite{Kingma2014Adam}.


\subsection{Baseline systems}

We use the BLSTM-based system and WPE-based system as our baselines.
\begin{itemize}
	\item WPE: For WPE-based system, we use the Matlab p-code\footnote{http://www.kecl.ntt.co.jp/icl/signal/wpe/} by the authors of \cite{nakatani2010speech,yoshioka2012generalization}.
	\item BLSTM: For BLSTM-based system, there are three BLSTM layers with 512 units, which keeps the network configuration the same as our proposed method. It also does the speech denoising and dereverberation simultaneously. Compared with our proposed method, the differences in the BLSTM-based system are that it does the speech denoising and dereverberation at one stage and without the deep embedding features.
\end{itemize}

\setlength{\tabcolsep}{2.0mm}{
	\begin{table*}[t]
		\caption{Different methods results on seen, unseen and average(AVG.) conditions. D is the dimension of embedding features.}
		\label{tab:results}
		\centering
		\begin{tabular}{c|c|ccc|ccc|ccc}
			\hline
			\multicolumn{1}{c|}{\multirow{2}*{Methods}}& \multicolumn{1}{c|}{\multirow{2}*{D}} & \multicolumn{3}{c|}{seen} & \multicolumn{3}{c|}{unseen} & \multicolumn{3}{c}{AVG.} \\
			\cline{3-11}
			&  & PESQ & CD(dB) & LLR(dB) & PESQ & CD(dB) & LLR(dB) & PESQ & CD(dB) & LLR(dB) \\
			\hline
			Unprocessed & -& 1.93 & 5.45 & 0.97& 2.05& 5.52& 0.88& 1.94& 5.46& 0.96\\
			\hline
			WPE(baseline) & -& 1.96& 5.40& 0.96& 2.06& 5.48& 0.87& 1.97& 5.41& 0.95\\
			BLSTM(baseline)& -& 2.61& 5.03& 0.81& 2.58& 4.92& 0.83& 2.60& 5.02& 0.81\\
			\hline
			& 10& 2.67& 4.48& 0.71& 2.62& 4.49& 0.71& 2.67& 4.48& 0.70\\
			Our& 20& 2.70& \textbf{4.39}& \textbf{0.68}& 2.64& \textbf{4.44}& 0.70& \textbf{2.70}& \textbf{4.40}& \textbf{0.68}\\
			proposed& 30& \textbf{2.71}& 4.42& \textbf{0.68}& 2.62& 4.47& \textbf{0.69}& \textbf{2.70}& 4.42& \textbf{0.68}\\
			& 40& \textbf{2.71}& 4.41& \textbf{0.68}& \textbf{2.65}& 4.46& 0.70& \textbf{2.70}& 4.42& \textbf{0.68}\\
			\hline
		\end{tabular}
\end{table*}}

\setlength{\tabcolsep}{2.5mm}{
	\begin{table*}[]
		\caption{The results of PESQ, CD(dB) and LLR(dB) for different methods on different SNRs.}
		\label{tab:results2}
		\centering
		\begin{tabular}{c|c|cccc|cccc|cccc}
			\hline
			&  D & \multicolumn{4}{c|}{{PESQ}}                                     & \multicolumn{4}{c|}{{CD(dB)}}                                   & \multicolumn{4}{c}{{LLR(dB)}}                                  \\ \hline
			SNR(dB)     & -& -5            & 0             & 5             & 10            & -5            & 0             & 5             & 10            & -5            & 0             & 5             & 10            \\ \hline
			Unprocessed & -& 1.53          & 1.85          & 2.05          & 2.35          & 6.33          & 5.60          & 5.36          & 4.53          & 1.32          & 0.98          & 0.91          & 0.63          \\ \hline
			WPE(baseline)         & -& 1.55          & 1.90          & 2.07          & 2.37          & 6.26          & 5.52          & 5.32          & 4.54          & 1.31          & 0.94          & 0.91          & 0.64          \\ 
			BLSTM(baseline)       & -& 2.17          & 2.57          & 2.75          & 2.92          & 6.12          & 5.03          & 4.73          & 4.18          & 1.14          & 0.82          & 0.74          & 0.55          \\ \hline
			& 10& 2.30          & 2.64          & 2.78          & 2.94          & 5.21          & 4.51          & 4.27          & 3.92          & 0.95          & 0.70          & 0.64          & 0.50          \\ 
			Our & 20& 2.33          & \textbf{2.67} & 2.82          & \textbf{2.97} & \textbf{5.09} & \textbf{4.41} & \textbf{4.21} & \textbf{3.88} & 0.93          & \textbf{0.68} & \textbf{0.63} & \textbf{0.49} \\ 
			proposed & 30&\textbf{2.34} & \textbf{2.67} & 2.82          & \textbf{2.97} & 5.11          & 4.44          & 4.23          & 3.91          & 0.93          & \textbf{0.68} & 0.64          & \textbf{0.49} \\ 
			 & 40& \textbf{2.34} & \textbf{2.67} & \textbf{2.83} & \textbf{2.97} & 5.12          & 4.43          & \textbf{4.21} & 3.90          & \textbf{0.92} & \textbf{0.68} & \textbf{0.63} & \textbf{0.49} \\ \hline
		\end{tabular}
	\end{table*}
}

\subsection{Evaluation metric}

In this work, in order to evaluate the quality of the enhanced speech, we compute the following objective measures. The perceptual evaluation of speech quality (PESQ) \cite{rix2002perceptual}, the Cepstral Distance (CD) \cite{kitawaki1988objective} and log likelihood ratio (LLR) \cite{hu2008evaluation} measures.

\subsection{Experimental results}

Table~\ref{tab:results} shows the results of PESQ, CD (dB) and LLR (dB) for different methods on seen, unseen and average(AVG.) conditions. The seen condition is for the 100 Nonspeech Sounds noise database, the unseen condition is the NISEX-92. Table~\ref{tab:results2} shows the results of PESQ, CD(dB) and LLR(dB) for different SNRs.

\subsubsection{The effectiveness of our proposed method}



From Table~\ref{tab:results}, we can find that our proposed speech dereverberation methods are superior to WPE and BLSTM baselines in all objective measures no matter seen or unseen condition. Compared with the BLSTM baseline, the proposed method produces a relative improvement of 3.8\% for PESQ measure, and a relative descending of 12.3\% and 19.1\% for CD and LLR measures. In addition, from Table~\ref{tab:results2} we can know that the performance of our proposed methods outperform the WPE and BLSTM baselines for all of the SNRs, especially in the low SNR condition. These results indicate the effectiveness of our proposed method. The reason is that our proposed method is a two-stage algorithm with joint training to optimize the enhanced model. Firstly, the noisy spectrums are mapped to noise-free vectors (deep embedding features), which is the speech denoising stage. Secondly, the dereverberation stage learns the target mask from these vectors. Finally, joint training is applied to optimize these two stages. Besides, the deep embedding features are more easily removed the reverberation than the amplitude spectral features. Because they contain the potential mask of anechoic speech and they are discriminative features. Therefore, they are conducive to speech dereverberation so that they can improve the performance of speech dereverberation.


\subsubsection{The effect of different embedding dimensions}

In order to acquire better deep embedding features, we select four different numbers of embedding dimension D (10, 20, 30 and 40) in this study. From Table~\ref{tab:results} and Table~\ref{tab:results2}, we can make several observations. First, different embedding dimensions have different performances of speech dereverberation, but they all have similar performance. They all show a better performance than baselines, which proves the robustness of our proposed method. Second, when \(D=20\), the proposed method gets the best performance in most of cases. However, as for the \(D=10\), it shows a slightly decreased performance no matter seen and unseen conditions. This indicates that too low dimension of the deep embedding features will damage speech dereverberation performance. This is because that low dimension can't represent the discriminative relation between the anechoic speech and reverberation very well. Therefore, the higher dimension of the deep embedding features can reduce the reverberation better. However, the dimension should not be very high, as we can see that when \(D=30\) or 40, performance gets slightly worse than \(D=20\) in some cases. The reason is that the higher dimension has more expensive computational cost, which may damage the performance of speech dereverberation. 

\section{Conclusions}

In this paper, we propose a joint training method for speech denoising and dereverberation using deep embedding features. The proposed method includes two stages: denoising and dereverberation. At the denoising stage, a DC network is trained to extract deep embedding features that are noise-free vectors. At the dereverberation stage, it directly learns the target mask form these deep embedding features. Finally, these two stages are optimized by the joint training method. Results show that our proposed method outperforms to WPE and BLSTM baselines. Compared to the BLSTM-based method, the proposed method produces a relative improvement of 3.8\% for PESQ measure, and a relative descending of 12.3\% and 19.1\% for CD and LLR measures. In the future, we will explore the proposed method for multi-channel speech dereverberation to make full use of the spatial information.

\section{Acknowledgements}

This work is supported by the National Key Research \& Development Plan of China (No.2017YFC0820602) and the National Natural Science Foundation of China (NSFC) (No.61425017, No.61831022, No.61773379, No.61771472), and Inria-CAS Joint Research Project (No.173211KYSB20170061).

\vfill\pagebreak

\bibliographystyle{IEEEtran}

\bibliography{mybib}

\begin{thebibliography}{10}
\providecommand{\url}[1]{#1}
\csname url@samestyle\endcsname
\providecommand{\newblock}{\relax}
\providecommand{\bibinfo}[2]{#2}
\providecommand{\BIBentrySTDinterwordspacing}{\spaceskip=0pt\relax}
\providecommand{\BIBentryALTinterwordstretchfactor}{4}
\providecommand{\BIBentryALTinterwordspacing}{\spaceskip=\fontdimen2\font plus
\BIBentryALTinterwordstretchfactor\fontdimen3\font minus
  \fontdimen4\font\relax}
\providecommand{\BIBforeignlanguage}[2]{{%
\expandafter\ifx\csname l@#1\endcsname\relax
\typeout{** WARNING: IEEEtran.bst: No hyphenation pattern has been}%
\typeout{** loaded for the language `#1'. Using the pattern for}%
\typeout{** the default language instead.}%
\else
\language=\csname l@#1\endcsname
\fi
#2}}
\providecommand{\BIBdecl}{\relax}
\BIBdecl

\bibitem{helfer1990hearing}
K.~S. Helfer and L.~A. Wilber, ``Hearing loss, aging, and speech perception in
  reverberation and noise,'' \emph{Journal of Speech, Language, and Hearing
  Research}, vol.~33, no.~1, pp. 149--155, 1990.

\bibitem{beutelmann2006prediction}
R.~Beutelmann and T.~Brand, ``Prediction of speech intelligibility in spatial
  noise and reverberation for normal-hearing and hearing-impaired listeners,''
  \emph{The Journal of the Acoustical Society of America}, vol. 120, no.~1, pp.
  331--342, 2006.

\bibitem{warzybok2014subjective}
A.~Warzybok, I.~Kodrasi, J.~O. Jungmann, E.~Habets, T.~Gerkmann, A.~Mertins,
  S.~Doclo, B.~Kollmeier, and S.~Goetze, ``Subjective speech quality and speech
  intelligibility evaluation of single-channel dereverberation algorithms,'' in
  \emph{2014 14th International Workshop on Acoustic Signal Enhancement
  (IWAENC)}.\hskip 1em plus 0.5em minus 0.4em\relax IEEE, 2014, pp. 332--336.

\bibitem{fan2019noise}
C.~Fan, B.~Liu, J.~Tao, J.~Yi, Z.~Wen, and Y.~Bai, ``Noise prior knowledge
  learning for speech enhancement via gated convolutional generative
  adversarial network,'' in \emph{2019 Asia-Pacific Signal and Information
  Processing Association Annual Summit and Conference (APSIPA)}.\hskip 1em plus
  0.5em minus 0.4em\relax IEEE, 2019, pp. 662--666.

\bibitem{kodrasi2018joint}
I.~Kodrasi and S.~Doclo, ``Joint late reverberation and noise power spectral
  density estimation in a spatially homogeneous noise field,'' in \emph{2018
  IEEE International Conference on Acoustics, Speech and Signal Processing
  (ICASSP)}.\hskip 1em plus 0.5em minus 0.4em\relax IEEE, 2018, pp. 441--445.

\bibitem{nakatani2010speech}
T.~Nakatani, T.~Yoshioka, K.~Kinoshita, M.~Miyoshi, and B.-H. Juang, ``Speech
  dereverberation based on variance-normalized delayed linear prediction,''
  \emph{IEEE Transactions on Audio, Speech, and Language Processing}, vol.~18,
  no.~7, pp. 1717--1731, 2010.

\bibitem{yoshioka2012generalization}
T.~Yoshioka and T.~Nakatani, ``Generalization of multi-channel linear
  prediction methods for blind mimo impulse response shortening,'' \emph{IEEE
  Transactions on Audio, Speech, and Language Processing}, vol.~20, no.~10, pp.
  2707--2720, 2012.

\bibitem{han2015learning}
K.~Han, Y.~Wang, D.~Wang, W.~S. Woods, I.~Merks, and T.~Zhang, ``Learning
  spectral mapping for speech dereverberation and denoising,'' \emph{IEEE/ACM
  Transactions on Audio, Speech, and Language Processing}, vol.~23, no.~6, pp.
  982--992, 2015.

\bibitem{wu2017reverberation}
B.~Wu, K.~Li, M.~Yang, and C.-H. Lee, ``A reverberation-time-aware approach to
  speech dereverberation based on deep neural networks,'' \emph{IEEE/ACM
  transactions on audio, speech, and language processing}, vol.~25, no.~1, pp.
  102--111, 2017.

\bibitem{williamson2017time}
D.~S. Williamson and D.~Wang, ``Time-frequency masking in the complex domain
  for speech dereverberation and denoising,'' \emph{IEEE/ACM transactions on
  audio, speech, and language processing}, vol.~25, no.~7, pp. 1492--1501,
  2017.

\bibitem{zhao2018late}
Y.~Zhao, D.~Wang, B.~Xu, and T.~Zhang, ``Late reverberation suppression using
  recurrent neural networks with long short-term memory,'' in \emph{2018 IEEE
  International Conference on Acoustics, Speech and Signal Processing
  (ICASSP)}.\hskip 1em plus 0.5em minus 0.4em\relax IEEE, 2018, pp. 5434--5438.

\bibitem{wang2014training}
Y.~Wang, A.~Narayanan, and D.~Wang, ``On training targets for supervised speech
  separation,'' \emph{IEEE/ACM transactions on audio, speech, and language
  processing}, vol.~22, no.~12, pp. 1849--1858, 2014.

\bibitem{fan2019discriminitive}
C.~Fan, B.~Liu, J.~Tao, J.~Yi, and Z.~Wen, ``Discriminative learning for
  monaural speech separation using deep embedding features,'' \emph{Proc.
  Interspeech 2019}, pp. 4599--4603, 2019.

\bibitem{fan2020end}
C.~Fan, J.~Tao, B.~Liu, J.~Yi, Z.~Wen, and X.~Liu, ``End-to-end post-filter for
  speech separation with deep attention fusion features,'' \emph{IEEE/ACM
  Transactions on Audio, Speech, and Language Processing}, 2020.

\bibitem{fan2020deep}
------, ``Deep attention fusion feature for speech separation with end-to-end
  post-filter method,'' \emph{arXiv preprint arXiv:2002.01626}, 2020.

\bibitem{Hershey2016Deep}
J.~R. Hershey, Z.~Chen, J.~L. Roux, and S.~Watanabe, ``Deep clustering:
  Discriminative embeddings for segmentation and separation,'' in \emph{IEEE
  International Conference on Acoustics, Speech and Signal Processing}, 2016,
  pp. 31--35.

\bibitem{fan2018Utterance}
C.~Fan, B.~Liu, J.~Tao, Z.~Wen, J.~Yi, and Y.~Bai, ``Utterance-level
  permutation invariant training with discriminative learning for single
  channel speech separation,'' in \emph{2018 11th International Symposium on
  Chinese Spoken Language Processing (ISCSLP)}.\hskip 1em plus 0.5em minus
  0.4em\relax IEEE, 2018, pp. 26--30.

\bibitem{wang2018supervised}
D.~Wang and J.~Chen, ``Supervised speech separation based on deep learning: an
  overview,'' \emph{IEEE/ACM Transactions on Audio, Speech, and Language
  Processing}, 2018.

\bibitem{fan2020spatial}
C.~Fan, B.~Liu, J.~Tao, J.~Yi, and Z.~Wen, ``Spatial and spectral deep
  attention fusion for multi-channel speech separation using deep embedding
  features,'' \emph{arXiv preprint arXiv:2002.01626}, 2020.

\bibitem{garofolo1993darpa}
J.~S. Garofolo, L.~F. Lamel, W.~M. Fisher, J.~G. Fiscus, and D.~S. Pallett,
  ``Darpa timit acoustic-phonetic continous speech corpus cd-rom. nist speech
  disc 1-1.1,'' \emph{NASA STI/Recon technical report n}, vol.~93, 1993.

\bibitem{kodrasi2018single}
I.~Kodrasi and H.~Bourlard, ``Single-channel late reverberation power spectral
  density estimation using denoising autoencoders,'' \emph{Proc. Interspeech
  2018}, pp. 1319--1323, 2018.

\bibitem{RIR}
E.~A.~P. Habets, ``Room impulse response (rir) generator,'' Available online:
  https://www.audiolabs-erlangen.de/fau/professor/habets/software/rir-generator
  (accessed on 21 March 2019), Tech. Rep.

\bibitem{100nonspeech}
G.~Hu, ``100 nonspeech sounds 2006 [oneline],'' Technical Report. Available
  online: http://web.cse.ohio-state.edu/pnl/corpus/HuNonspeech/HuCorpus.html
  (accessed on 22 February 2019), Tech. Rep.

\bibitem{varga1993assessment}
A.~Varga and H.~J. Steeneken, ``Assessment for automatic speech recognition:
  Ii. noisex-92: A database and an experiment to study the effect of additive
  noise on speech recognition systems,'' \emph{Speech communication}, vol.~12,
  no.~3, pp. 247--251, 1993.

\bibitem{Abadi2016TensorFlow}
M.~Abadi, A.~Agarwal, P.~Barham, E.~Brevdo, Z.~Chen, C.~Citro, G.~S. Corrado,
  A.~Davis, J.~Dean, and M.~Devin, ``Tensorflow: Large-scale machine learning
  on heterogeneous distributed systems,'' 2016.

\bibitem{Kingma2014Adam}
D.~P. Kingma and J.~Ba, ``Adam: A method for stochastic optimization,''
  \emph{Computer Science}, 2014.

\bibitem{rix2002perceptual}
A.~W. Rix, M.~P. Hollier, A.~P. Hekstra, and J.~G. Beerends, ``Perceptual
  evaluation of speech quality (pesq) the new itu standard for end-to-end
  speech quality assessment part i--time-delay compensation,'' \emph{Journal of
  the Audio Engineering Society}, vol.~50, no.~10, pp. 755--764, 2002.

\bibitem{kitawaki1988objective}
N.~Kitawaki, H.~Nagabuchi, and K.~Itoh, ``Objective quality evaluation for
  low-bit-rate speech coding systems,'' \emph{IEEE Journal on Selected Areas in
  Communications}, vol.~6, no.~2, pp. 242--248, 1988.

\bibitem{hu2008evaluation}
Y.~Hu and P.~C. Loizou, ``Evaluation of objective quality measures for speech
  enhancement,'' \emph{IEEE Transactions on audio, speech, and language
  processing}, vol.~16, no.~1, pp. 229--238, 2008.

\end{thebibliography}


\end{document}